# White and green rust chimneys accumulate RNA in a ferruginous chemical garden

Vanessa Helmbrecht[1], Maximilian Weingart[2], Frieder Klein[4], Dieter Braun[2], William D. Orsi[1,3,*]

[1]Department for Geo- and Environmental Sciences, Palaeontology & Geobiology, Ludwig-Maximilians-Universität, 80333 Munich, Germany

[2]Systems Biophysics, Faculty of Physics, Ludwig-Maximilians-Universität, 80799 Munich, Germany

[3]GeoBio-Center[LMU], Ludwig-Maximilians-Universität München, 80333 Munich, Germany

[4]Department of Marine Chemistry and Geochemistry, Woods Hole Oceanographic Institution, Woods Hole, MA 02543, USA

*Corresponding author: w.orsi@lrz.uni-muenchen.de

**Abstract**

Mechanisms of nucleic acid accumulation were likely critical to life's emergence in the ferruginous oceans of the early Earth. How exactly prebiotic geological settings accumulated nucleic acids from dilute aqueous solutions, is poorly understood. As a possible solution to this concentration problem, we simulated the conditions of prebiotic low-temperature alkaline hydrothermal vents in co-precipitation experiments to investigate the potential of ferruginous chemical gardens to accumulate nucleic acids via sorption. The injection of an alkaline solution into an artificial ferruginous solution under anoxic conditions ($O_2$ <0.01% of present atmospheric levels) and at ambient temperatures, caused the precipitation of amakinite ("white rust"), which quickly converted to chloride-containing fougerite ("green rust"). RNA was only extractable from the ferruginous solution in the presence of a phosphate buffer, suggesting RNA in solution was bound to $Fe^{2+}$ ions. During chimney formation, this iron-bound RNA rapidly accumulated in the white and green rust chimney structure, as it was depleted from the surrounding solution. Our findings reveal that in the oceans of the early Earth, white and green rust chimneys were likely key geochemical features that can rapidly sequester and accumulate RNA. This represents a new mechanism for nucleic acid accumulation, in addition to wet dry cycles, and may have promoted RNA survival in a dilute prebiotic ocean.

**Introduction**

There are two main theories on the emergence of life on Earth, the "RNA world" theory[1–5] and the "metabolism first" in alkaline hydrothermal vents (AHV) theory[6–10]. Both theories are controversial and often discussed as mutually exclusive, but is this really the case? There have been many disagreements between the representatives of these theories, leading to decade-old divisions in the origins of life research[11–13]. Few studies have shown evidence for compatibility of an RNA world within an AHV environment[14,15]. For instance, there is some evidence that RNA can oligomerize in AHVs[16,17]. A key unknown however, is how nucleic acids could overcome the large dilution factor of prebiotic oceans, which has been referred to as the "concentration problem"[18–20]. AHVs with the ability to accumulate nucleic acids could offer a plausible reaction site and solve the problem of RNA dilution in a prebiotic ocean.

The famous Lost City hydrothermal field near the Mid-Atlantic Ridge is probably among the best studied AHVs today[21,22]. In opposition to acidic high-temperature hydrothermal vents, Lost City offers conditions, that might have been favorable for life's emergence. The serpentinization-dominated hydrothermal system vents alkaline fluids at ambient temperatures, co-precipitating mainly calcium carbonates and brucite upon mixing with seawater[21]. However, compared to modern oceans, prebiotic oceans of the Hadean and Archean were likely more acidic and ferruginous, i.e. anoxic conditions with high concentrations of dissolved iron[23–27]. The acidity of prebiotic seawater[6,28] probably prevented the crystallization of acid-soluble minerals, like the characteristic minerals of Lost City[21,22]. Therefore, Lost City type chimneys were most certainly not present on early Earth.

Despite significant differences in the oceanic composition between modern and prebiotic oceans, serpentinization was an ongoing process. As soon as the hydrosphere stabilized ~4.2 Ga ago[29,30] (Figure 1A), water would have interacted with ultramafic to mafic seafloor rocks, resulting in widespread hydrothermal alteration, including serpentinization across the Hadean seafloor[31,32] (Figure 1B). Thus, serpentinization has featured prominently in the AHV theory for the emergence of life, as it creates strongly reducing conditions that favor the abiotic synthesis of organic compounds[8]. As seawater seeped into faults and fractures in the young oceanic lithosphere, chemical exchange between the percolating fluid and the igneous rocks at low temperatures would likely have changed its composition and resulted in more alkaline fluids. The pH of the emanating fluids is largely temperature dependent, with circumneutral pH at high temperatures (~300°C) and increasingly alkaline pH with decreasing temperatures[31,33].

To summarize, the likely ferruginous conditions of the prebiotic oceans suggest the existence of a different kind of AHV on early Earth (Figure 1C). Previous studies argue for the presence of "green rust" as an important mineral in the emergence of life at AHVs, since green rust can form inorganic membranes to maintain disequilibria between the acidic ocean and the alkaline interior of the vents[17,34–38]. The highly reactive mineral green rust is a ferrous-ferric oxyhydroxide also known as fougerite[36,39,40]. Amakinite or "white rust", the rare ferrous analogue of brucite, is the precursor phase of fougerite and was detected in the basement rocks below the Lost City hydrothermal field[41,42].

Despite acknowledging the minerals' significance in an emergence of life context, experimental evidence for alkaline vents made out of white and green rust chimneys, is scarce[17,43]. Moreover, their capability to accumulate nucleic acids to avoid dilution in a prebiotic ocean, has not been tested so far. Therefore, this study focuses entirely on the formation of white to green rust transforming chimneys (Figure 1C) and their ability to bind and accumulate RNA, offering a potential solution to the concentration problem by De Duve[18]. We created alkaline vents under ferruginous conditions ($O_2$ <0.01% of present atmospheric levels). Within these chemical gardens, chimney structures composed of the iron-containing hydroxide white rust and the layered double hydroxide green rust form across strong pH gradients. We show that the chimneys accumulate RNA from the surrounding ferruginous solution. In a prebiotic ocean on early Earth, this process may have helped to overcome dilution of the RNA to very low concentrations.

**Methods**

Experimental setup

Similar to previous AHV experiments[17,35,44,45], our experimental setup was designed to mimic ferruginous early Earth conditions (Figure 1B). Three AHV experiments were performed in an anaerobic chamber (MBRAUN Labstar) containing a $N_2$ atmosphere to maintain anoxic conditions. Oxygen values were continuously monitored inside the anaerobic chamber using two digital sensors. They were constantly below <0.01% of present atmospheric levels. All AHV experiments were carried out at room temperature (25°C) as it seems likely that low-temperature systems were widespread in off-axis environments.

Our experimental setup was similar to those used in previous studies where an alkaline solution representing the hydrothermal fluid is slowly pumped into a ferruginous solution representing the ocean[16,35,44–48] (Supplemental Figure 1). We used an inverted 125ml glass vial with the bottom cut off to hold the ferruginous solution. The vial was sealed with a rubber stopper (Supplemental Figure 1) and a pipette tip was inserted through the crimp-seal and connected to a polymer tube on the outside (bottom) of the inverted flask, which was connected to a syringe 10ml Luer-lock plastic syringe (B Braun) filled with alkaline NaOH (Carl Roth) solution. The polymer tubing was connected to the syringe via a Sterican® hypodermic-needle (B Braun, 1.20x50mm, 18G x 2). The syringe was mounted on a single-syringe pump (Cole Parmer, Model 100 78-9100C), that was vertically oriented to ensure a constant flow rate of alkaline fluid from the syringe into the simulated iron ocean. Horizontal orientation of the syringe and syringe pump was found to result in less stable chimneys, because the required longer tubing length caused introduction of gas bubbles that disrupted chimney growth. Vertical orientation of the syringe and syringe pump, and the shorter tubing length greatly ameliorated this problem.

A 0.2M solution of $FeCl_2 \cdot 4H_2O$ (Sigma-Aldrich) and MilliQ water resulted in ferruginous and acidic (pH 3) conditions. The alkaline (pH 13) hydrothermal solution consisted of 0.2M NaOH (Carl Roth) dissolved in MilliQ water. All solutions were prepared inside the anaerobic chamber to ensure anoxic conditions and reduce the presence of dissolved $O_2$ in the solutions. The alkaline NaOH solution was injected into the ferruginous solution at a rate of 3.7ml/h. The pH of the ocean solution was monitored over the duration of the experiment using a pH meter. After ~2 hours and injection of 7ml NaOH, chimney growth ceased. Dry-powdered total RNA from the yeast *Saccharomyces cerevisiae* (Sigma-Aldrich, catalog number 10109223001) was dissolved in the ferruginous solution with an initial concentration of 0.12 mg/ml.

RNA sampling, extraction and quantification

Inside the anaerobic chamber, 1ml samples of the ferruginous solution were taken with a pipettor at set time intervals as the chimney grew. At the end of the experiment, the chimney pieces were dried in the anaerobic chamber for 24 hours and frozen for subsequent RNA extraction (see below). All RNA extractions from the iron ocean and the green rust chimneys were performed in a laminar flow clean bench equipped with a high-efficiency particulate air filter. The clean bench is a dedicated bench for RNA only with its own dedicated set of pipettors, racks and tubes, where no DNA samples or PCR products are introduced[49,50]. This helps to maintain a relatively low level of RNA and DNA contamination from aerosols as well as DNA contamination from PCR products. All surfaces are sterilized with UV prior and after sample processing and all surfaces are wiped down with RNAseZap to remove any potential contaminating microbes or aerosolized RNA.

During the experiment within the anaerobic chamber, ferruginous solution was sampled at a consistent frequency (every minute to every 10 minutes) to capture a high temporal resolution from the initial period of chimney formation where the most rapid RNA depletion in the ferruginous solution simulant was observed. The ocean samples were transferred to RNA/DNA clean 2ml Eppendorf tubes that were frozen at -20°C. RNA was extracted by adding 1ml of acidic (pH: 4.0) trizol (MP Biomedicals) and 1M $Na_2HPO_4$ (Sigma-Aldrich) powder. The addition of the phosphate buffer was used to reduce RNA adsorption to iron particles in the ocean solution, which is a well-known feature of iron minerals that can reduce nucleic acid extraction efficiencies[51]. The solutions were mixed by vortexing for 10 seconds and centrifuged for 10 minutes at 4°C (to increase RNA preservation and strengthen the trizol phase separation) at 13000rpm. The supernatant (containing the dissolved RNA) was transferred via pipetting to a new 2ml RNA/DNA free tube. An equal volume of ≥99.5% isopropanol (Carl Roth) and 3µl glycogen (Carl Roth) as a carrier were added to help precipitate the relatively low concentrations of RNA. Precipitation was done overnight at -20°C. RNA was pelleted by centrifugation (5 minutes at 4°C, 13000rpm). The pellets were air dried in an RNA-dedicated laminar flow hood (see above) with HEPA filtered air to reduce contamination. To remove residual salts, the pellets were washed with 500µl of 50% molecular biology grade ethanol (Carl Roth) centrifuged for 5 minutes and air dried in the laminar flow clean hood. After air drying, the RNA pellets were resuspended in RNAse free diethyl pyrocarbonate (DEPC) treated water (Carl Roth) and vortexed for 30 seconds to resuspend the RNA. Quantification was done with an Invitrogen Qubit 3 Fluorometer and the micro–RNA kit.

After the AHV experiment, the ferruginous solution was drained with a pipette and the chimney was dried out for one day inside the anaerobic chamber. The dried chimney was broken into separate pieces as replicates for RNA extractions, weighed, and placed in sterile RNA free 2ml screw cap microtubes (MP Biomedicals, Lysing Matrix E) containing 0.5ml of differently sized silica beads. The silica beads were added to the RNA extraction step to physically disrupt the chimney structure during beating in order to enhance RNA retrieval during the extraction. RNA extractions were optimized using 1ml of acidic (pH: 4) Phenol:Chloroform:Isoamyl alcohol ('trizol'). To desorb the RNA off of the green rust, 1M $Na_2HPO_4$ was added since the addition of a phosphate buffer assists in desorbing RNA off of mineral surfaces[51]. Chimney samples were homogenized in the presence of the acidic trizol and phosphate buffer in the Lysing Matrix E tubes, using a Fastprep-24 5G (MP Biomedicals) at a speed of 6.0m/sec for 40 seconds. The samples were centrifuged for 10 minutes at 4°C and 13000rpm and the upper aqueous phase (containing the RNA) was transferred to new 2ml Eppendorf tubes. The RNA was precipitated by adding an equal volume of isopropanol and 3µl glycogen each. All samples were frozen overnight at -20°C. Pelleting, washing and measuring the RNA was done as described in the previous paragraph.

Raman spectroscopy

The dry green rust chimneys were stored under an $N_2$ atmosphere to minimize oxidation prior to the analysis with the Raman spectrometer (HORIBA Jobin Yvon XploRA, Mineralogical State Collection Munich). For data acquisition, the chimneys were taken out of the anaerobic chamber. Exposure to air was less than 2 minutes until the start of the first Raman measurement.

After 30 minutes, the chimney was completely oxidized and no green rust was detectable anymore. The settings of the scans were as follows: spectral range from 50 to 4000cm$^{-1}$, 10 seconds acquisition time, 2 accumulations, 300µm confocal hole and 100µm slit, 1800 grooves/mm grating and a filter with 10 to 25% optical density of the incoming laser to avoid laser-induced oxidation of the chimney. A green laser with a wavelength of 532nm, the x100 long working distance (LWD) objective (0.80 numerical aperture) were used for all measurements and the Labspec 6 software was used to process the acquired spectra. We used an in-house reference Raman database MSC-RD (Mineralogical State Collection Raman Database[52]), the KnowItAll database (Horiba Edition, now John Wiley & Sons) and the database from the RRUFF$^{TM}$ Project[53]. The instrument was calibrated with a silicon wafer (520.7cm$^{-1}$) daily and immediately before the measurements took place.

Scanning electron microscopy

We used a Phenom XL G2 scanning electron microscope (SEM) to further characterize the mineralogy of chimneys under vacuum. The dry chimneys were stored under an $N_2$ atmosphere until measurements started. Air exposure during transfer to the vacuum chamber of the SEM was limited to 30 seconds maximum. All measurements were carried out using an acceleration voltage of 10kV, a beam current of 1µA-pA, a vacuum pressure of 0.10Pa and a beam diameter of 0.005 to 1µm. For imaging, secondary electrons (SE) and backscattered electrons (BSE) were used. Elemental analysis of the green rust chimneys was carried out with the electron dispersive X-ray spectroscopy detector (EDX).

Imaging, processing, quantifying chimney growth

Images of chimneys were taken with a Sony Alpha a6000 mirrorless camera and a 90mm Sony macro lens. The images were processed using the Adobe Lightroom and Photoshop software packages and the height of the chimney was determined in Adobe Photoshop with a ruler. Timelapse videos were taken with the Sony Alpha a6000 as well. The camera took an image of the chimney every 2 seconds during growth, which are 600 images in 20 minutes. The raw timelapse video from the camera was sped up again in post-production to ~3.3 minutes per second (link to full video: first 40 minutes of growth in 12 seconds).

**Results**

Chimney Morphology

The AHV experiments were carried out in an anaerobic chamber in a $N_2$ atmosphere. During chimney growth, no pH changes were observed in the ferruginous solution. In all three experiments, the injection of the alkaline fluid resulted in the formation of chimney structures within 60 minutes. These structures resemble "chemical gardens" that have been found to form in AHV models across steep pH gradients[17]. The average (+/- standard deviation) height of the chimneys was (3.5 +/- 0.5) cm. The average width was (0.6 +/- 0.2) cm. The fastest growth rates were observed in the first 10 minutes of the experiments (Figure 3). After the initial growth phase, which ended after ~20 minutes, vertical chimney growth slowed down significantly (Figure 3), after which horizontal growth of small mm-sized branches was often observed. No more vertical growth was observed after 50 to 60 minutes. Growth of additional chimney extensions was detected only in 1 out of 3 experiments after one hour of chimney growth.

A reproducible color transition pattern was observed during all experiments, whereby the base of the chimney would change from being translucent or white color to a green color in five minutes only. Then, as the chimneys grew, the green color gradually moved upwards towards the top replacing the white color by the end of the experiment (Supplemental Video 1). After a few minutes, the surface of the green precipitate would break and a new chimney structure would grow vertically on top of the green structure (Supplemental Video 1). These new structures also appeared first as a white film, which turned green also within five minutes of exposure to the ferruginous solution (Supplemental Video 1). This process of constant ruptures, re-precipitation and color transition continued, until the chimneys reached their maximum heights. A visualization of the phase transition from the white film to green colored chimney structure is presented in Figure 3E and in Supplemental Video 1.

Green Rust Identification

The fully grown chimney shows a complex morphology on a macroscopic and a microscopic scale. SEM images show that the chimney is composed of a thin inorganic 'membrane' of euhedral hexagonal plate-shaped crystals (Figure 2A) that are similar in size and shape to green rust[38,54]. Crystals formed aggregates, with individual crystals ranging in diameter from < 1μm to approximately 20μm. The chimneys contain an inorganic membrane with a width between 80 to 90μm (Figure 2A). SEM images also revealed a high internal pore space and surface area with nanopores (<1μm) and channels in the chimney structure (Figure 2A).

Raman spectra of a dried chimney suggests that it is mainly composed of green rust (Figure 2C). This evidence is based on the characteristic Raman bands at 427 and 518cm$^{-1}$, that are considered diagnostic for green rust identification[40,54]. The EDX spectrum highlights the elemental composition of the chimneys, which is consistent with chloride-bearing green rust (Table 1 in Figure 2B). Chloride green rust has a chemical composition of ~$[Fe_3^{2+}Fe^{3+}(OH)_8]^+[Cl \cdot 2H_2O]^-$ [34,35,55]. When left exposed to air for more than 20 minutes, the chimneys changed color quickly from dark green to an ochre color indicating oxidation of iron. Based on SEM and Raman analyses, these oxidation products were identified as hematite (Figure 2A, Supplemental Figure 2).

RNA depletion from the iron ocean during chimney growth

A rapid decrease of RNA in the ferruginous solution was observed primarily during initial stages of chimney growth, particularly in the first 10-30 minutes when the chimney is white and transitions to green color (Figure 3B). This was confirmed in a second experiment, where RNA was depleted to levels below detection in the simulated ocean within the first 10 minutes (Figure 3D). As a negative control, this trend of rapid RNA depletion was not observed in the absence of an alkaline injection, hence no chimney structure (Figure 3A). The rate of natural RNA degradation in this blank experiment under the ferruginous conditions (without a chimney present) was relatively slow and only decreased <10% over the course of the experiment, which was 3.5 hours. (Figure 3A). In contrast, in the presence of a white and green rust chimney structure, that forms in the presence of alkaline venting, the RNA was depleted ≥99% and was below the detection limit of the Qubit Fluorometer by the end of the experiment (Figure 3B). No RNA could be extracted from the ferruginous solution without the addition of a phosphate (1M $Na_2HPO_4$) buffer (Figure 3C), indicating that the dissolved RNA is likely adsorbed onto iron particles[51] in the "iron ocean".

There is an inverse correlation ($R^2$ = 0.92) between the size of the chimney and the concentration of RNA in the iron ocean (Supplemental Figure 4). The relationship is described by an exponential function, whereby as the chimney increases in height at a linear rate of growth, RNA concentrations in the ferruginous solution are depleted at an exponential rate (Supplemental Figure 3).

RNA accumulation in chimneys

When a chimney was present, RNA concentrations decreased 1,000-fold in the ferruginous solution compared to the starting concentration (Figure 4A). This contrasts with conditions of a blank experiment, where no alkaline solution was injected, hence no chimney was formed. In this negative control, only a minor decrease in RNA concentration in the iron ocean was observed (Figure 3A). At the end of the experiments containing a green rust chimney, concentration of RNA in the chimney (893 ± 63 ng per gram) was significantly higher (two-sided t-test: P = 0.0003) than the concentration of RNA in the ferruginous solution at the end of the experiment (35 ± 11 ng per gram) (Figure 4A). Without a phosphate extraction buffer, no RNA was extractable from the chimney. This indicates that the RNA was complexed with the iron in the chimney, a well-known feature of iron containing minerals (and clay minerals in general) that bind nucleic acids and reduce their extractability[51].

**Discussion**

Experimental geochemical conditions in the context of predictions for early Earth

We tested key geochemical aspects of the prebiotic ocean on early Earth, anoxic and ferruginous conditions. It is widely accepted that these seawater conditions were characteristic of the Hadean and Archean[23–27]. Our alkaline vent experiments are based on the prediction, that widespread serpentinization of the oceanic lithosphere was common in prebiotic oceans[37,56]. This probably resulted in venting of high pH fluids at the seafloor, forming alkaline springs when the alkaline fluids came in contact with the ferruginous ocean.

Our experiments were carried out at 25°C room temperature, which seems justified considering that temperature estimates of the prebiotic ocean are highly uncertain[17,56], ranging from boiling to freezing conditions[57–59]. Hydrothermal activity can proceed over an even broader range[60] from <10°C to about 407°C. Thus, our ambient experimental conditions represent a relatively cool off-ridge environment. We also show, that hydrothermal conditions with elevated temperatures are not necessarily required for the formation of green rust chimneys in a ferruginous solution. Relatively mild temperatures are not out of the question in terms of prebiotic ocean conditions, since recent models[61] predict temperatures for some regions of the prebiotic ocean down to as low as 0 to 50°C. The fact that our green rust chimneys form at 25°C is also beneficial to RNA stability compared to elevated temperatures[62–64].

We were not able to simulate high pressures in the anaerobic chamber. Thus, all experiments were performed at atmospheric pressure, which is relatively low and not representative of the deep sea, where many hydrothermal vents are found[65]. However, shallow water alkaline vents might have been located near oceanic islands, island arcs or continental shelves, if they existed in the Hadean and Archean, which is highly debatable[66]. Nonetheless, many of the earliest fossils known today are located in shallow water settings, like the 3.4 Ga Strelley Pool Formation in the Pilbara Craton of Western Australia[67]. Moreover, modern examples of such shallow serpentinite-hosted vents today are the Strytan Hydrothermal Field in Iceland or the Prony Hydrothermal Field in New Caledonia[68], showing the possibility of hosting complex biological communities. In conclusion, water depth and pressure mimicked in our experiments could represent shallow water vents, that hold some of the earliest signs of life in the fossil record[67], but the results need to be reviewed with caution.

Assessing oxidation of the chimneys

It is known that green rust oxidizes quite fast upon air exposure[54]. Therefore, we carefully assessed the possibility for oxidation in our experiments. Oxygen exposure during the alkaline vent experiments ($O_2$ <0.01% of present atmospheric values) and sample transfer (transport in $N_2$ purged gas tight flasks) was kept at a minimum. We acknowledge, that some oxidation might have occurred during the transfer of the chimneys to the microscope. The SEM and elemental analysis indicate effects of oxidation to be minor and resulted in small hematite globules (<10 µm in size) that are likely the first signs of oxidation (Figure 2A). This likely has to do with SEM measurements being made in a vacuum chamber with no air, and the sample being exposed to air for only a few seconds during transfer. However, when the sample was scanned with the Raman it was not in a vacuum and showed strong signs of oxidation after 20 minutes where the entire chimney had been oxidized to hematite (Supplemental Figure 2). When the green rust chimney was measured with the Raman within one or two minutes, the characteristic Raman bands at 419 and 506 $cm^{-1}$ for green rust[54] were observed (Figure 2C). Therefore, our results indicate that we are forming green rust in the anaerobic chamber that quickly oxidizes to hematite under the Raman. However, it is possible to obtain the green rust spectra with the Raman when measuring quickly after transfer from the gas tight anoxic flask. As expected, the oxidation in air changes the green rust to hematite relatively quickly.

Mineralogy and morphology of the chimneys

Evidence for injection-growth experiments under ferruginous conditions forming white and green rust, is scarce. For this reason, we simulated iron hydroxide chimneys in an anaerobic chamber. The first white phase that precipitates in our alkaline vent experiments is probably $Fe(OH)_2$ sometimes referred to as amakinite[42]. It shows the typical white color for ferrous iron hydroxide[69] and turns green when it comes in contact with the ferruginous solution in less than five minutes (Figure 3), indicating oxidation and green rust formation[35,40]. Difficulties arose, when we attempted to investigate the white rust phase, because of its quick transformation into green rust during the AHV experiment and the lacking ability to dry this highly metastable phase in order to microscopically analyse it. Similarities in other studies point to this first precipitated phase being ferrous hydroxide $Fe(OH)_2$. Stone et al.[70] also performed chemical garden chimney growth experiments across pH gradients in ferruginous solutions, and identified the oxidized precipitate of iron in aqueous solutions as ferrous hydroxide $Fe(OH)_2$, namely "white rust". Trolard et al.[42]

describe the precipitation of ferrous hydroxide after mixing an Fe(II) salt solution with sodium hydroxide. Barge et al.[35] identify the first precipitates in a green rust forming experiment as white rust (iron hydroxide) as well. White rust forms by oxidation and partial hydrolysis of the dissolved iron[40], which is the process described in our experiments.

Green rust showed a higher stability than white rust under anoxic conditions, which allowed us to dry the chimneys and prepare them for measurements. Since $FeCl_2*4H_2O$ was used as a salt for the ferruginous solution, chloride-containing green rust GR1-Cl was most likely formed out of the precursor phase white rust[38,42]. Depending on the intercalated anion in the double layer hydroxide structure, different varieties of green rust are possible, $Cl^-$, $CO_3^{2-}$ or $SO_4^{2-}$ green rust for example[37,40]. The hexagonal and tabular morphologies of the crystals (Figure 2A) are a typical indicator for green rust as well[38,40,54]. Our chimneys show very similar Raman bands to previous studies about green rust[54,71], although they are slightly shifted to 419 and 506 cm$^{-1}$ (Figure 2C).

Chimney morphologies like ours (but with varying chemistries) have been created in different AHV experiments in the past[16,35,44,46–48,72]. The spontaneously formed chemical gardens are made out of vertically oriented chimneys with high internal surface areas that act as flow-through chemical reactors composed of inorganic membranes[47]. The formation mechanism consists of continuous bursting and branching[44], until a chemical garden precipitate is formed. The mainly vertical formation of the chimney structures might be due to fluid buoyancy effects, suggesting the injected sodium hydroxide was less dense than the iron-rich solution[73]. Experiments performed in microgravity showed no preferred chimney growth orientation[73–75]. The injection of 7ml of a pH 13 NaOH solution had no measurable effect on the pH of the ferruginous solution during the experiment. This could either indicate that the $OH^-$ was being consumed completely by the green rust chimney structure as it grew over time, or the amount of NaOH pumped into the ferruginous solution was simply too small to overcome its buffering capacity.

RNA-iron complexes in the ferruginous solution

A fundamental question answered in our AHV experiments is the timing of the RNA sequestration. Does the RNA bind to already precipitated white and green rust chimneys, or do RNA-iron complexes build the chimneys while they grow? The latter implies the formation of RNA-iron complexes in the ferruginous solution before the chimney formation. McGlynn et al.[76] support the theory of iron hydroxide (and iron sulfide) chimneys at alkaline vents as possible polymerization sites of RNA in life's emergence and also mention the complexation between RNA ligands and metal ions. Nonetheless, evidence for RNA-iron complexation and chimney formation under ferruginous conditions has poorly been documented in previous studies. The findings of our study suggest, that RNA is sequestered into the chimney structure by binding to the iron ions present in the solution first. After complexation, the RNA-iron complexes gradually build the chimney as it grows (Figure 4B). Hence, the iron acts as a transporter for the RNA into the chimney structure. This is evident, because no RNA was extractable from the ferruginous solution without a phosphate buffer[51,77] (Figure 3C).

Generally, adsorption of RNA onto mineral surfaces is a known process particularly in the origin of life context[17,51,77–80]. Moreover, the binding of RNA to ferrous iron to form complexes is described in previous studies[81,82]. The complexation likely occurs through ionic interactions between the positively charged iron ions and the negatively charged phosphate groups of the RNA[77]. In addition, a stronger effect of divalent cations, like $Mg^{2+}$ or $Fe^{2+}$, on nucleic acid adsorption onto mineral surfaces than monovalent ($Na^+$, $K^+$) cations has been identified[77]. Thus, the white and green rust chimneys, which are composed of lots of $Fe^{2+}$, present a favorable location for RNA to adsorb and accumulate. The sequence of nucleic acids doesn't have a notable effect on the adsorption strength[77].

Another significant aspect of green rust is, that it structurally resembles phyllosilicates with plate-shaped crystals (Figure 2A) and swelling capacities because of their ability to absorb water in the interlayer space. Green rust contains brucite-like layers of divalent and trivalent cations and interlayers of anions and structural water[54]. Due to their sheet-like nature, green rust is also called an anionic clay[54]. But not only the structural similarities are significant, when talking about green rust and clay minerals. Previous studies show, that phyllosilicates like montmorillonite can adsorb and oligomerize nucleic acids[83–85]. Other studies mention the successful adsorption of nucleotides onto Fe-Mg-rich swelling clays, for example nontronite[78,86]. Our findings indicate, that green rust's similar characteristics with clay minerals favor the accumulation of nucleic acids like RNA, particularly within AHV chimneys that form under our experimental conditions.

RNA accumulation dynamics in the green rust chimney

The correlation between the height of the chimney and the amount of RNA in the ferruginous solution indicates that the amount of RNA that accumulates in the chimney is dependent on how large the chimney structure is (Supplemental Figure 3). When the chimney height did not change, the amount of RNA also did not change. However, RNA accumulation inside the chimney could also be correlated to the phase transition. RNA seems to bind faster to the initial white rust phase compared to green rust that forms several minutes later (Supplemental Video 1). Most of the RNA has already been concentrated in the chimney in the first 10 minutes, when the chimney was not yet fully green (Figures 3B, 3D).

In the blank negative control experiments without a chimney, the natural RNA degradation (Figure 3A) was minimal, which does not explain the depletion of RNA in the ferruginous solution in the presence of a chimney. Instead, the RNA gets depleted rapidly if a chimney structure is present in the ferruginous solution (Figure 3B, 3D). Thus, the white and green rust chimney structure offers a favorable and highly reactive location for the RNA to accumulate (Figure 4A). This is evident when considering that at the end of the experiments we extracted higher concentrations of RNA from the dried green rust chimney compared to the surrounding solution (Figure 4A). Therefore, the RNA is moving from the ferruginous solution into the chimney over time, as it grows, and becomes concentrated in the green rust chimney.

Implications for Emergence of Life Theories

By reducing the variables down to a few key factors (ferruginous conditions, anoxia, and alkaline springs) we can study the effects of the white to green rust chimney transformation and its influence on RNA accumulation, since especially green rust has been hypothesized to be a key mineral for the emergence of life. Particularly Altair et al.[17] and Russell[37] highlight green rust as a key mineral in life's emergence, for example due to its redox reactivity, its ability to form inorganic membranes and to maintain ionic disequilibria, or its hydrous interlayers functioning as engines powered by these ionic disequilibria to drive essential endergonic reactions. This underlines the significance of green rust chimneys that form in our ferruginous chemical gardens.

As it has been demonstrated by our study, RNA-metal ion complexation and adsorption to mineral surfaces, most likely played a significant part during the emergence of life. Highly reactive mineral surfaces, especially at AHVs, can concentrate nucleic acids or nucleotides by adsorption, which saves them from dilution in the prebiotic ocean[78]. The effects of iron-complexed RNA should be considered in future RNA-world debates, and also how ribozymes bind or chelate to iron in a ferruginous solution. Ribozymes, which are enzymes, whose catalytic centers are entirely made of RNA[87], may be the next higher functional building block on the way to a living organism. Ribozymes, which can cleave, join and replicate, are key components for protein synthesis, since they are part of the ribosomes of cellular organisms[88]. Thus, it is believed that ribozymes preceded DNA and proteins during evolution[88]. Successful trapping and accumulation of ribozymes at iron hydroxide chimneys in future experiments would not only underline the hypothesis of white rust and green rust as important minerals, but of AHVs in general during the emergence of life. Furthermore, Furthermore, the presence of $Fe^{2+}$ over $Mg^{2+}$ in prebiotic oceans probably enhanced the catalytic activities of ribozymes[82].

We acknowledge, that because our tested RNA came from yeast, it is not an ideal proxy for ribozymes. Because the RNA is derived from a living organism (yeast), our experimental setup is not immediately comparable to a possible RNA-world that might have consisted of pre-biotic self-replicating RNAs. Our experimental setup is similar to that used by McGlynn et al.[76] who also used

RNA from yeast to test hypotheses surrounding the influence of iron sulfide and iron hydroxide minerals on RNA-complexation and binding dynamics in hydrothermal chimney structures. Using a similar logic, we used yeast RNA as a simplification to study how RNA accumulation can occur in white and green rust chimneys forming under simulated geochemical conditions of the early Earth (ferruginous, anoxic). Our results show that in principle, white and green rust chimney structures can bind and concentrate this RNA potentially providing a solution to the "concentration problem" for RNA in a prebiotic ocean. Despite deriving from a living organism, the yeast RNA shares many biochemical and biophysical properties with ribozymes. For example, yeast RNA and ribozymes should both represent soluble charged biomolecules that would promote binding iron minerals like green rust. However, the shorter strand length of ribozymes and potentially different or lower nucleotide diversity of the ribozymes could affect binding to the white rust and green rust, which should be a focus of future studies to see if the same concentration dynamics seen here for yeast RNA also applies to ribozymes.

**Outlook**

The mineral green rust has been proposed as one of the key minerals in life's emergence, which underlines the significance of our results that show, white to green rust transforming chimneys accumulate RNA from a ferruginous solution. No rocks exist on the Earth that are older than 4 billion years, but our experiments indicate that green rust chimneys may have been a feature of the early Earth seafloor at alkaline springs under ferruginous conditions after the formation of a stable hydrosphere around ~4.2 billion years ago[29]. RNA concentrating mechanisms may have promoted survival of self-replicating RNAs, that otherwise would be diluted to extremely low concentrations in the early ferruginous oceans during the Hadean and Archaean.

In contrast to alkaline vents, it was suggested that wet-dry cycles in surface pools[20] are a better place for concentrating biomolecules at life's emergence than submarine vents, given the "concentration problem" caused by the dilution effect of the ocean[20,89–91]. We want to emphasize with our study that white and green rust composed chimneys accumulate RNA from a ferruginous solution, thereby providing experimental evidence that AHVs should be considered in the emergence of life debate as potential accumulation sites for nucleic acids like RNA. In summary, our results provide a potential explanatory mechanism to solve the "concentration problem" by De Duve[18] for RNA in a prebiotic ocean.

**Acknowledgements**

We thank Dr. Dan Mills for valuable feedback on geobiological interpretations of the experiments. We acknowledge Ninos Hermis for assistance with preliminary experimental setups. This work was supported by the Deutsche Forschungsgemeinschaft (DFG, German Research Foundation)—Project-ID 364653263—TRR 235 to WDO and DB, and under Germany's Excellence Strategy—EXC 2077-390741603.

**Author contributions**

VH and WDO designed the experiments. VH performed all AHV experiments and RNA extractions. FK, MW, DB, WDO and VH analyzed the data. All authors commented on the manuscript and contributed to the writing process.

**Conflict of interest**

The authors declare no conflict of interest.

**Data availability**

Data sharing not applicable.

**Additional information**

Supplemental Figures 1-3 and Supplemental Video 1.

**Figures and figure legends**

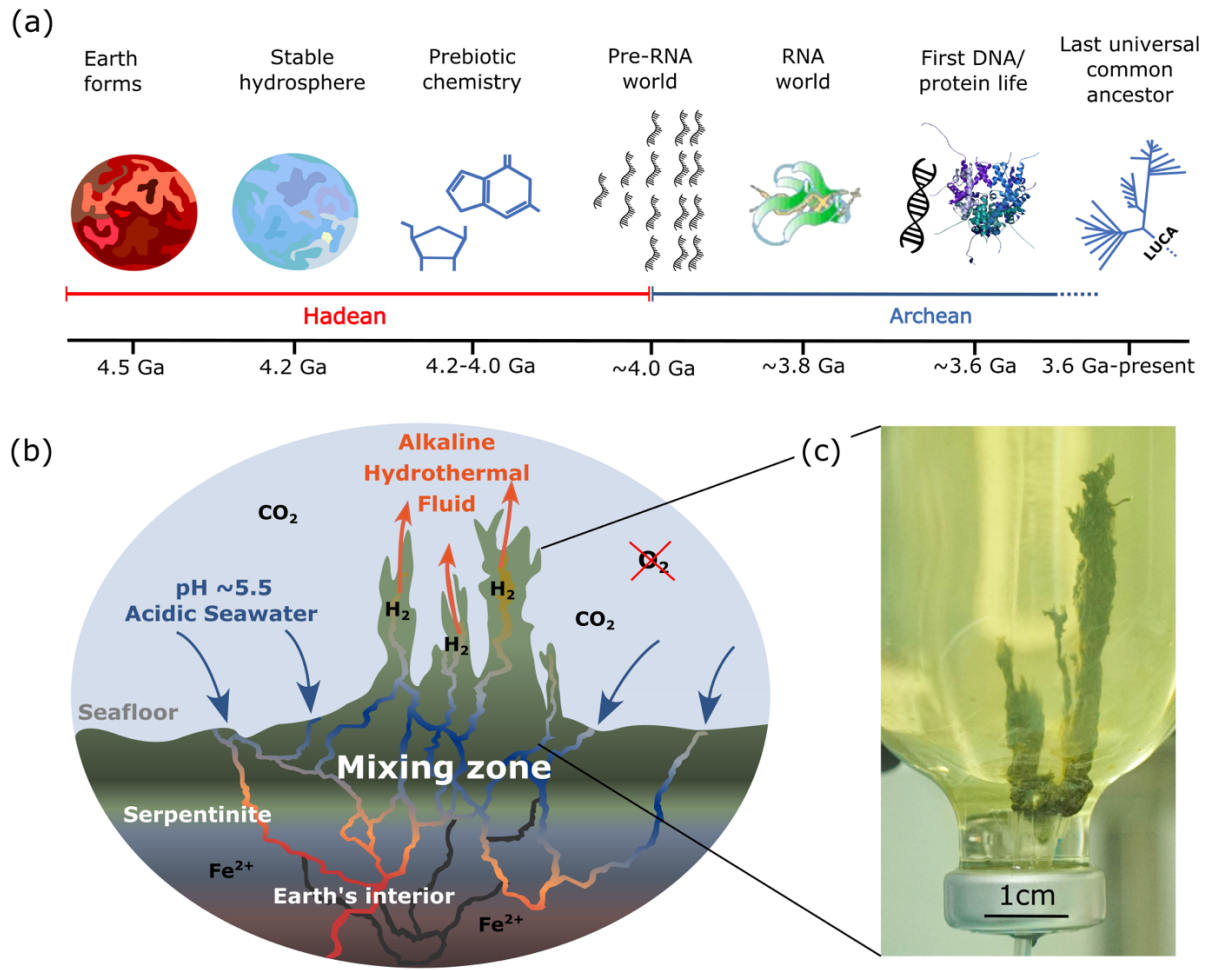

**Fig. 1 Alkaline vents and the emergence of life in a ferruginous environment.** (a) A hypothetical timeline of events possibly leading to the emergence of life on Earth, modified after Joyce[29]. (b) Schematic drawing of the seafloor conditions during the ferruginous oceans of the Hadean and Archean. The pH of the seawater was around 5.5. Acidic seawater gets reduced by $Fe^{2+}$ containing rocks in the Earth's interior, producing molecular hydrogen, which fuels the formation of alkaline vents. (c) Green rust chimney formed in the anaerobic chamber through the mixing of an alkaline fluid with a ferruginous solution.

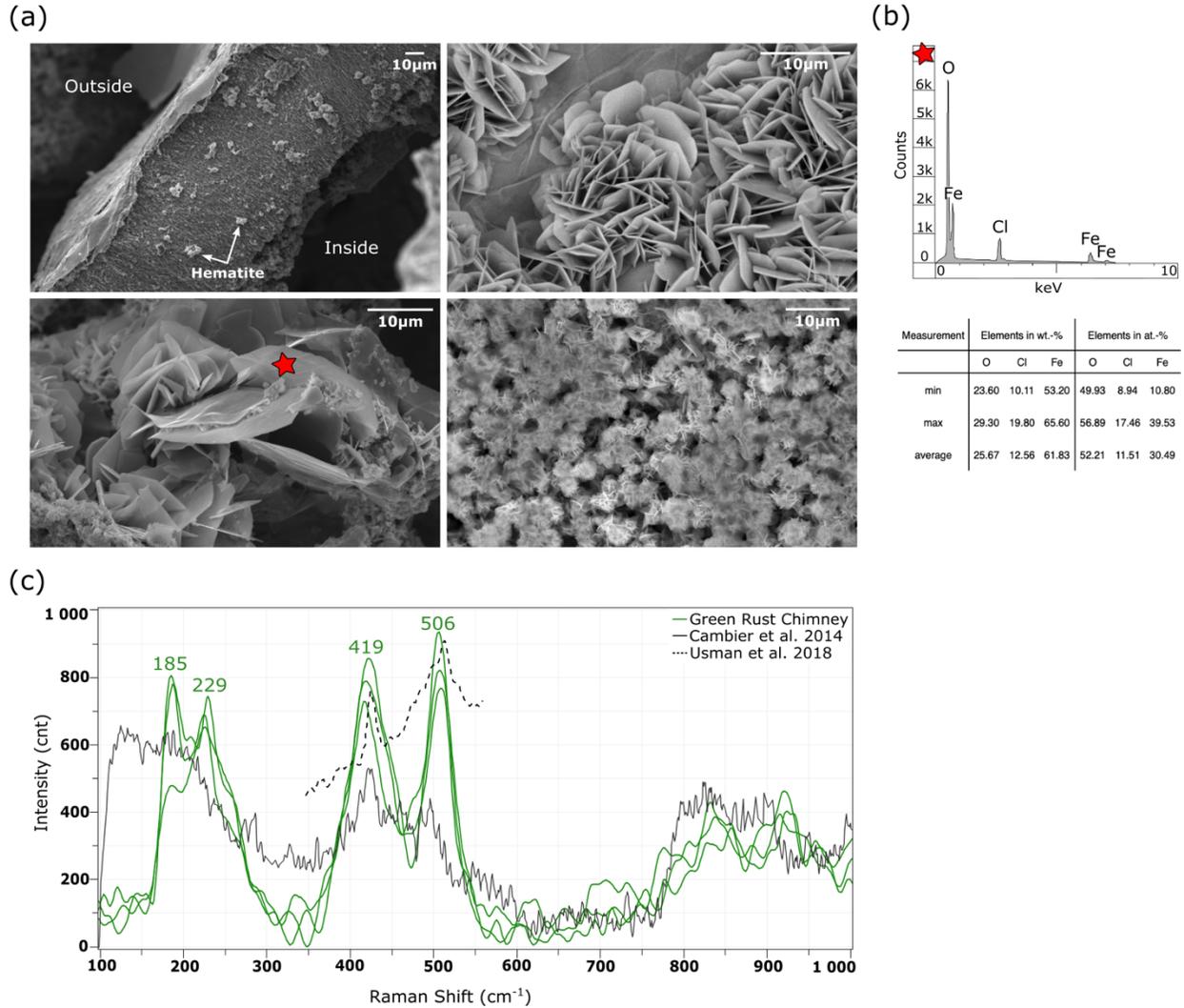

**Fig. 2 Dried chimney structures are composed primarily of green rust.** (a) BSE images of a dried green rust chimney. The first image shows a chimney wall with a thickness of approx. 80 to 90μm. Hematite globules are located on top of the green rust. The other BSE images show different magnifications of hexagonal platy green rust crystals. Contamination with oxygen during the transfer to the vacuum chamber was limited to max. 1 minute transfer time. (b) EDX spectrum with the elemental composition of the chimney in the marked region (red star). Acquisition settings were 30 seconds time, 10 kV acceleration voltage and 0.10 Pa vacuum pressure. Table 1 shows the average chemical composition of the chimney in five different spots. (c) Raman spectra of a green rust chimney made with 0.2M NaOH and 0.2M $FeCl_2*4H_2O$. The significant Raman bands for green rust identification from Cambier et al.[54] are at 427 and 518 cm$^{-1}$. Other mineral peaks except green rust peaks were removed from the Cambier et al.[71] spectrum for simplification.

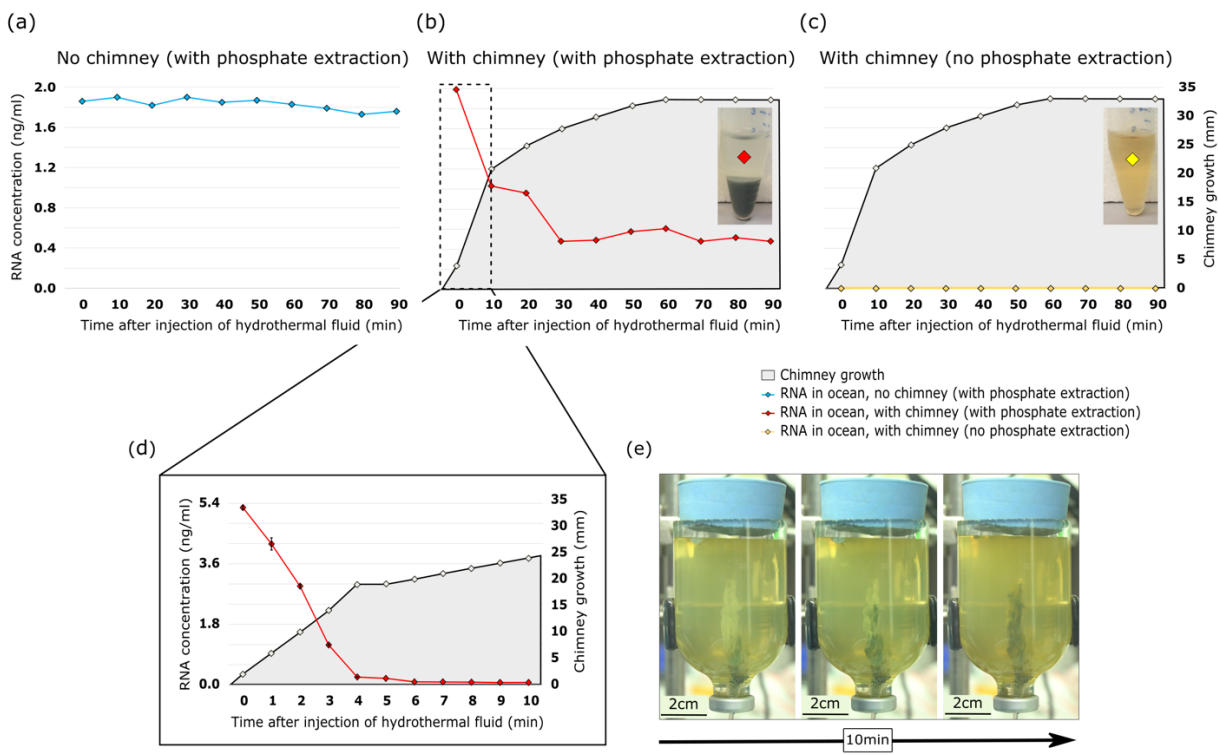

**Fig. 3 RNA is depleted from the ferruginous solution only in the presence of white and green rust chimneys**. (a) Rate of RNA depletion in the ocean solution without chimney growth (blank experiment) over a 90-minute time interval. RNA extraction was performed by adding phosphates to desorb the RNA from the iron in the simulated ocean solution. (b) Rate of RNA depletion from the ocean, and growth of the chimney structure (gray area) over a 90-minute time interval. The depletion of RNA is now significantly higher than in panel A with no chimney present. RNA extraction was performed by adding phosphates to desorb the RNA from the iron in the ocean solution. (c) RNA extracted without phosphates during chimney growth. This highlights the effect of adding phosphates to the RNA extraction protocol. When no phosphates were added, there was no extractable RNA from the ferruginous solution (yellow line). (d) The first 10 minutes of the experiment in panel B were sampled multiple times to give a higher resolution of the main area of interest, where chimney growth was the fastest. Error bars represent the standard error of the mean for RNA measured across three technical replicates (where error bars are not visible the standard error was <0.01). RNA extraction was performed with phosphates again. (e) White $Fe(OH)_2$ (possibly amakinite) oxidation to green rust in contact with the ferruginous solution. The duration of the color change was 10 min.

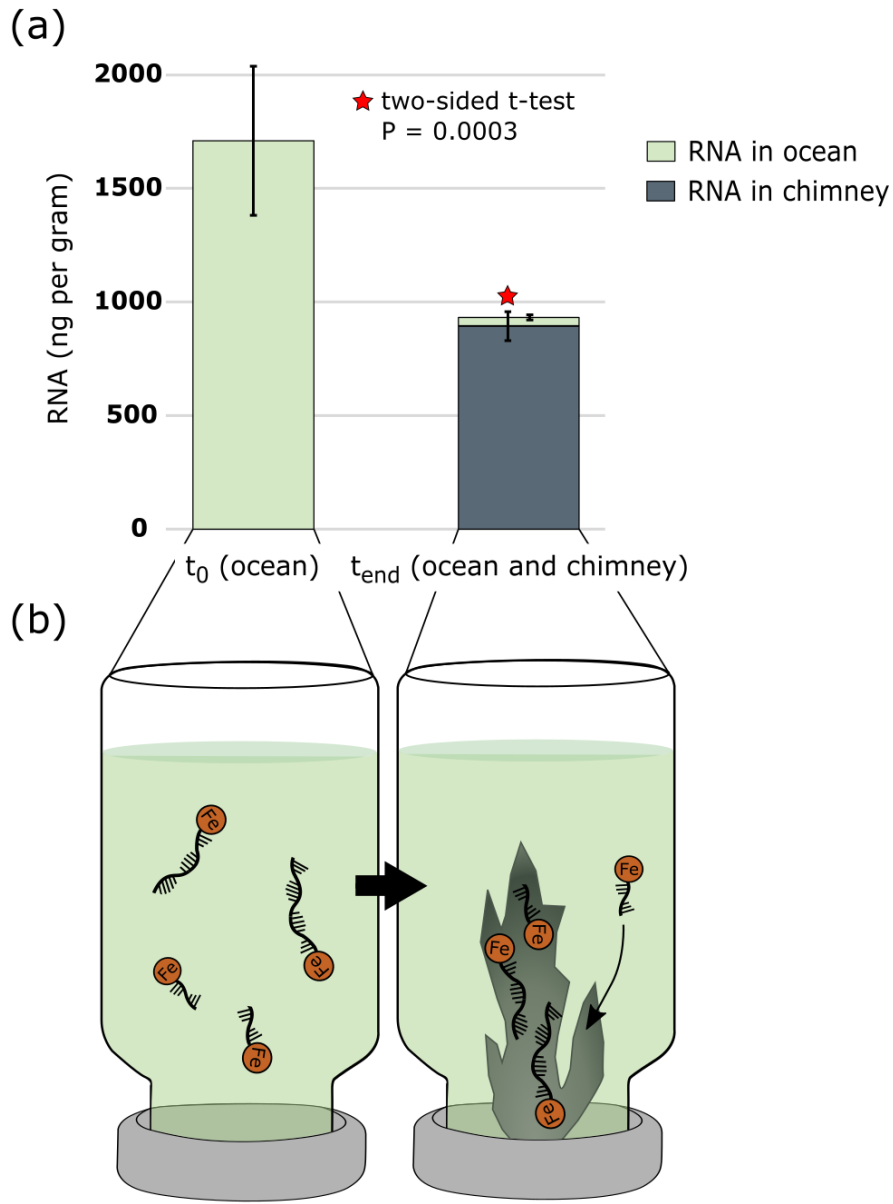

**Fig. 4 Accumulation of RNA in the white and green rust chimney structure.** (a) Bar charts show the concentration of RNA in the ferruginous solution at the beginning ($t_0$) and the end of the experiment ($t_{end}$) (green bar charts), as well as the retrieved RNA from the dried chimneys (grey bar chart). Given are the average values from three different experiments. Error bars represent standard error of the means. The two-sided t-test was significant with P=0.0003. (b) Schematic drawing of Fe-RNA complexes in the ocean solution getting incorporated into the chimney structure while it grows. At $t_{end}$, no RNA was detectable in the ocean solution, as it was only detectable in the chimney (panel A).